\documentclass{article}
\usepackage{amsmath}
\usepackage{geometry}%
\usepackage{caption}%
\usepackage{color}
\usepackage{float}%
\usepackage{amsfonts}%
\usepackage{amssymb}%
\usepackage{graphicx}
\usepackage{booktabs}
\usepackage{siunitx}
\providecommand{\U}[1]{\protect\rule{.1in}{.1in}}

\begin{document}
\[%
\begin{tabular}
[c]{c}%
$\underset{}{}${\LARGE Impact of COVID-19 Lockdowns on House Sparrows:}$\underset{}{}$\\
$\underset{}{}${\LARGE Comparative Study from an Indian Context}$\underset{}{}$\\
\end{tabular}
\]

\[%
\begin{tabular}
[c]{c}%
\text{Sudeep R. Bapat}$^{1,*}$ and \text{Sreeranjini T. M.}$^{2}$ \\ \\
$^1$Indian Institute of Management Indore\\ 
$^2$University of Hyderabad\\ \\
$^*$sudeepb@iimidr.ac.in
\end{tabular}
\]%

\noindent\textbf{Abstract: }Due to the outbreak of COVID-19, the last couple of years have been drastic in terms of human behavioural patterns. People were forced to stay at home for a very long duration because of the strict lockdown measures imposed by governments all over the globe. India was no exception, wherein the Indian government imposed several very strict lockdowns all across the country, which restricted human activities and their social behaviours. However, such restrictions were seen to have a positive impact on environment and ecology. In this paper, we aim to study the changes in House Sparrow sightings, as a result of the lockdowns. Are the house sparrows back? is the question we try to answer, using appropriate exploratory analysis and statistical modelling.

\vspace{0.05in}\vspace{0.05in}


\bigskip

\section{Introduction}

\noindent The ``House sparrow", scientifically known as \textit{Passer domesticus} is one of the most commonly found birds across India. Since a very long time, the relationship between human beings and sparrows has been harmonious. They are easily observable in windows and balconies of houses, which shows how well they have adapted with the human lifestyle. A typical house sparrow sighted in India is as seen in Figure 1. But since the starting of the 21st century, their population has been threatened and the population curve is also dipping. Several reasons have been cited for this decline, one of them being the increasing urbanization and pollution. Organizations like Nature Forever Society (NFS) have adopted several measures to conserve sparrows like ``Common Bird Monitoring of India", ``World Sparrow Day", ``Project Save Our Sparrows", among others. In literature, there have been many research articles published in recent times, which also suggest a decline in the number of house sparrows. Some of these include, an article by Sharma and Binner (2020), where the authors indeed claim a decline in their number, according to a study conducted by the Indian Council of Agricultural Research (ICAR). Even though the cause of decline was not known specifically, it is believed that it was due to the non-availability of nests, due to modernization and urbanization. Another article is that by Paul (2015), where similar conclusions are drawn. Narayanappa (2022) conjectured that the house sparrow is on the verge of decline due to urbanization which leads to the unavailability of suitable nesting sites. Modak (2017) studied the impact of urbanization on house sparrow distribution, particularly in the city of Kolkata, India. A few other related articles which one may refer to are by Ghosh et al. (2010), Chaudhary (2020) or Deepalakshmi and Salomi (2019). 

In this paper, we focus on a related issue, that of studying the effects of COVID-19 lockdowns on house sparrow counts. As discussed earlier, the pandemic impacted the entire world drastically. Human behaviours and social interactions were hampered, mainly because of the strict lockdowns imposed by governments across the globe. However, such lockdowns were a boon to the ecology and the environment. The wildlife sightings increased after COVID hit and especially during lockdowns. One belief is that since the humans were not allowed to roam much, they could spend more time observing and recording such sightings. There have already been a lot of literature stating that the lockdowns indeed impacted the wildlife in a positive way. One may refer to Basile et al. (2021), where the authors investigated how stay-at-home orders affected data submitted by birdwatchers in Italy, Spain and the United Kingdom. Schrimpf et al. (2021) found that counts of several focal bird species changed in pandemic-altered areas, usually increasing in comparison to prepandemic era. Seress et al. (2021) assessed changes
in reproductive success of great tits \textit{(Parus major)} at two urban habitats, due to the COVID pandemic. Gordo et al. (2021) studied the birds’ response to the population lockdown by using bird records collected by a citizen science project in northeastern Spain. 

\begin{center}
\includegraphics[width = .45\textwidth]{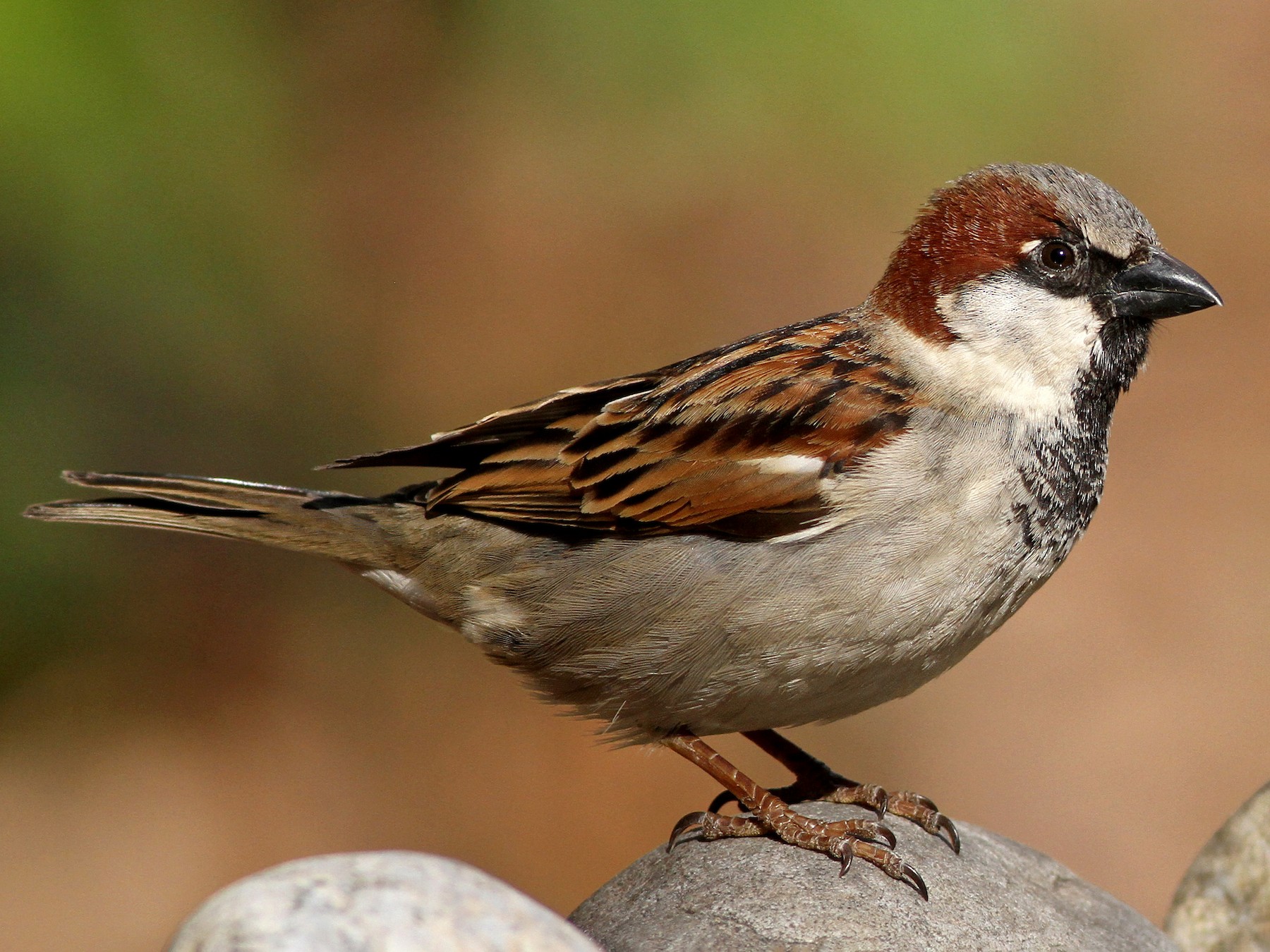}\\
\textbf{Figure 1}. Image provided by eBird (www.ebird.org) and created on [16/04/2015]
\end{center}

The structure of the paper is as follows: Section 2 provides the methodology used for the analysis, along with a brief description of the dataset. Section 3 covers the statistical analysis and modelling of the sparrow counts using appropriate count regression models, while Section 4 includes a brief discussion.

\section{Methodology}

\noindent In this paper, we aim to study the effect of the COVID-19 pandemic and hence the reduction in human noise on the count of house sparrows in the country of India. In order to do so, we downloaded data on house sparrow count in India from \textit{e-bird}, which is one of the largest biodiversity science projects with more than 100 million bird sightings contributed annually by eBirders around the world.
On January 30th, 2020, India’s first ever COVID case was reported in the state of Kerala. As the number of COVID cases increased, the Prime Minister of India announced a nationwide lockdown on the midnight of March 24th, which  lasted initially for 21 days. Further, it was extended up to May 31st, 2020. The movement and activities of the citizens got restricted as a result of this imposition of lockdown. This sudden pause in the human activities also astonishingly brought down the air and noise pollution levels. This gave us a rare opportunity to carry out a comparative study about how the less crowded and noisy cities affected the lives of birds like house sparrows. One of the assumptions is that the strict ``stay at home" orders might have brought more people into birdwatching and the time spent on such activities might have increased. Thus, we hypothesized that there could be an increase in the daily number of observations after the imposition of lockdown.  

For this study and analysis, our dataset contains 187898 \textit{e-bird} house sparrow checklists from 01/01/2018 to 30/11/2021 collected by bird observers across India. We tried to analyze the changes in the number of bird counts during lockdown as compared to that in previous years. We did a yearly comparison of the daily number of counts from 2018 to 2021, which very well includes the COVID phase as well. Additionally, we performed a phase-wise comparison during the COVID period, by focusing on the daily bird count in each phase, where the phases are formed by dividing the timeline to `before lockdown', `during lockdown' and `after lockdown'. For a better understanding, the following points outline the phases which we chose, along with the corresponding dates.
\begin{itemize}
\item Phase 1: before first lockdown (before 24/03/2020)
\item Phase 2: during first lockdown (24/03/2020 to 31/05/2020)
\item Phase 3: before second lockdown (01/06/2020 to 04/04/2021)
\item Phase 4: during second lockdown (05/04/2021 to 15/06/2021)
\item Phase 5: after second lockdown (16/06/2021 to 31/11/2021)
\end{itemize}
As mentioned before, the checklists had a total of 187898 observations across 29 variables which include observation count, county, state, locality (hotspot or personal), date of sighting, time of sighting, duration, number of observers in the group etc. Firstly, as we wanted to focus on the daily bird counts, we aggregated the different entries within each day to a single number. Another aspect of study was to assess the trends in sparrow counts around sunrise (4 am to 7 am) and sunset (5 pm to 7 pm). The belief was that these two time periods differ from the others in terms of bird sightings, as the sparrows are more active and less harmed by pollution and human intrusion. Finally, we try to model the sparrow counts using an appropriate Poisson regression model to conclude that there was indeed an effect of the lockdowns on their count. Further results and analysis are presented in the subsequent sections.

\section{Statistical Analysis and Modelling}

This section provides a wide array of exploratory and statistical techniques applied on different sets of bird counts, as a comparison and modelling approach. Before we begin with any detailed analysis, we will provide a glimpse of the spatial distribution of sparrows in India, according to the different states. Figure 2 contains side-by-side gradient maps of India showing the sparrow counts before COVID (before 30/01/2020) and after. One can note the clear increase in sparrow counts in the same states after the outbreak of COVID. 

\[%
\begin{tabular*}
{\textwidth}[c]{@{\extracolsep{\fill}}cc}%
$\includegraphics[width = .4\textwidth]{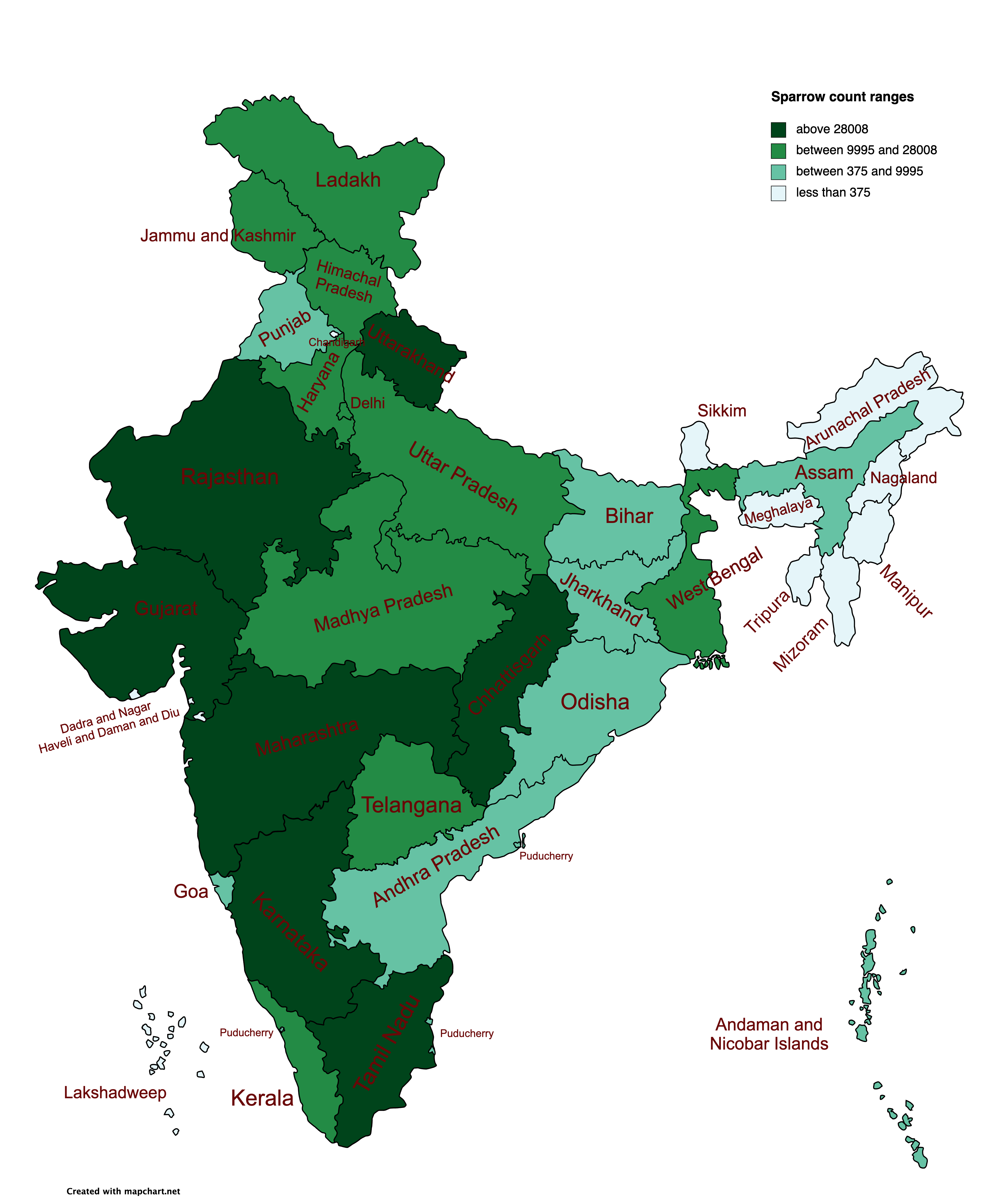}$ & $\includegraphics[width = .4\textwidth]{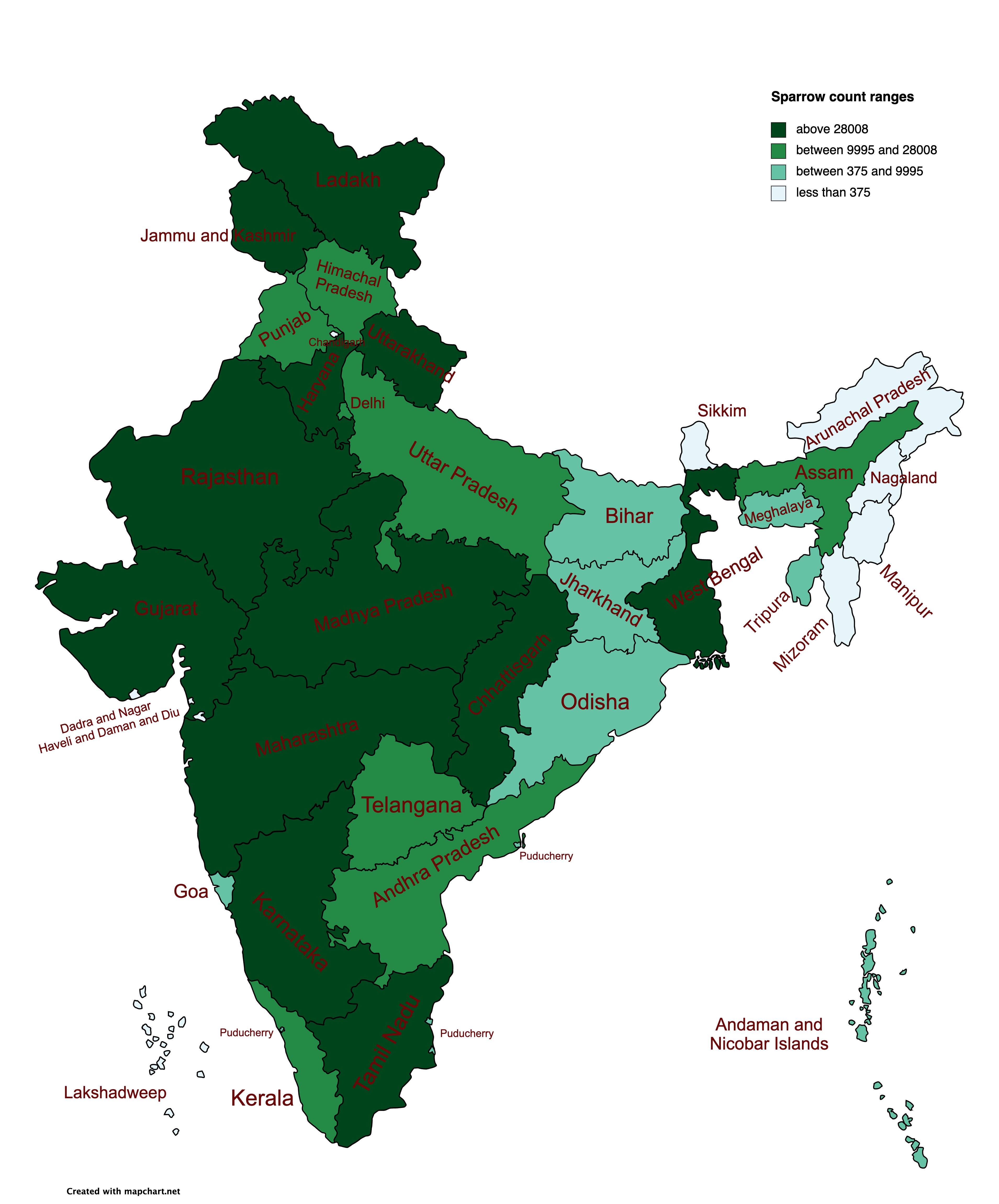}$\\
\multicolumn{2}{c}{\textbf{Figure 2}. Side-by-side maps of India showing the sparow counts before (left) and after (right) COVID}
\end{tabular*}
\]

\subsection{Year wise comparison}

The first aspect is to conduct an appropriate year wise comparison of bird counts. To begin with, we wanted to assess the trends in daily observation counts over time. We hence aggregated the daily observation count for each date from 1st Jan 2018 to 30th Nov 2021, into a single number. Basic boxplots and histograms revealed the presence of outliers in the data, which were removed, separately for each year, using the standard approach of judging the interquartile range (if an observation is 1.5 times the interquartile range more than the third quartile (Q3) or 1.5 times the interquartile range less than the first quartile (Q1), it is considered an outlier). Figure 3 contains side-by-side boxplots, histograms and separate scatterplots showing the distribution of bird counts for each year from 2018 to 2021. Right away, one can notice the increase in the count in the years 2020 and 2021, as compared to the previous ones. Especially the histograms show that the counts become more symmetric for years 2020 and 2021. This gives a visual affirmation of the fact that sparrows were sighted more during COVID as compared to before.

\[%
\begin{tabular*}
{\textwidth}[c]{@{\extracolsep{\fill}}cc}%
$\includegraphics[width = .45\textwidth]{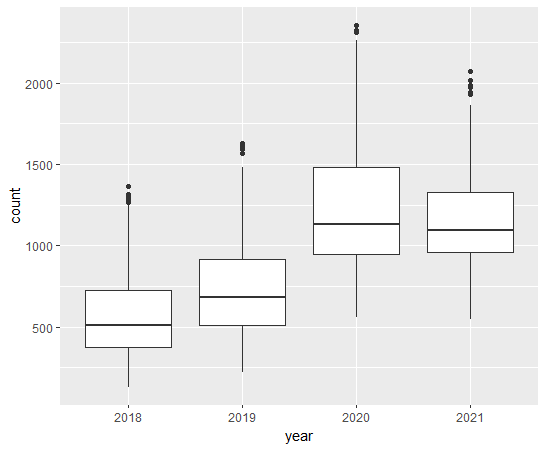}$ & $\includegraphics[width = .45\textwidth]{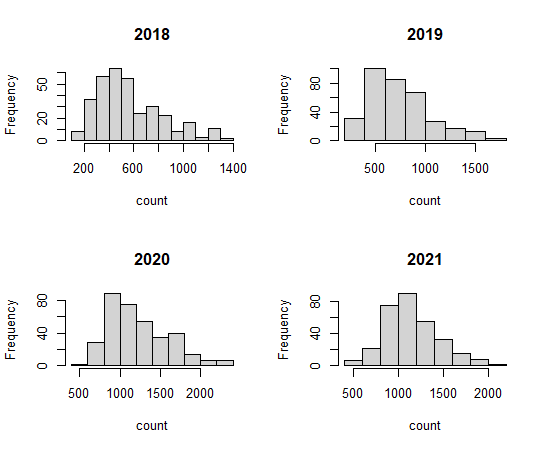}$
\end{tabular*}
\]

\begin{center}
\includegraphics[width = .45\textwidth]{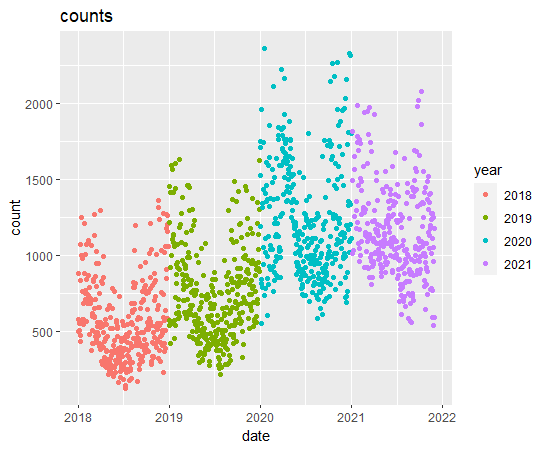}\\
\textbf{Figure 3}. Side-by-side boxplots (top left), histograms (top right) and scatterplots (bottom) for the bird counts
\end{center}

\bigskip

\[%
\begin{tabular*}
{\textwidth}[c]{@{\extracolsep{\fill}}c|ccccccc}%
\multicolumn{8}{c}{\textbf{Table 1}. Summary variables of bird counts for each year}\\ \\  \hline
Year & No. of days & Min & $Q_1$ & Median & Mean & $Q_3$ & Max \\\hline \\
$2018$ & $336$ & $131$ & $372$ & $508$ & $566.5$ & $723.5$ & $1365$  \\
$2019$ & $344$ & $222$ & $508$ & $684$ & $740.2$ & $913.2$ & $1629$  \\
$2020$ & $349$ & $558$ & $944$ & $1130$ & $1224$ & $1481$ & $2356$  \\
$2021$ & $305$ & $544$ & $956$ & $1092$ & $1149$ & $1325$ & $2075$ \\ \hline \\
\end{tabular*}
\]

\bigskip

Basic summary variables of the bird counts for each year is also presented in Table 1 for completeness. One can clearly note that the average bird count has indeed increased after the outbreak of COVID.

Further, we also conducted a few other comparisons between the bird counts, by focusing on different time periods. The first comparison was between `before COVID' and `after COVID' sparrow counts. In order to do so, we considered the data from 01/03/2019 to 28/02/2020 as `before COVID' and paired it with the same dates next year namely, 01/03/2020 to 28/02/2021, which was termed as `after COVID'. A suitable way to conduct this comparison is through a non-parametric approach, namely the Wilcoxon signed rank test. This is an ideal way to compare the locations of two matched samples. The hypotheses to test here are whether the true location shift equals zero or is less than zero. The corresponding p-value which was obtained was $< 2.2 \times 10^{-16}$, and we were able to conclude that the counts `after COVID' were indeed substantially more.

Another possible comparison is that by looking at the entire available data before 30/1/2020, when the first COVID case was discovered in India, and that after 30/1/2020. Since both the samples can be considered to be independent, we applied another non-parametric approach namely, the Wilcoxon rank sum test, This test is capable of finding whether the locations for both the independent samples are equal or not. The obtained p-value was again $< 2.2 \times 10^{-16}$, which concludes that the number of sparrow sightings increased after the very initial COVID outbreak in India. A more visual representation of this conclusion can be seen from the comparative histograms given in Figure 4. The y-axis gives the frequencies of the corresponding counts. One can easily note that for the `after' group, higher frequencies correspond to larger bird counts.

\begin{center}
\includegraphics[width = .55\textwidth]{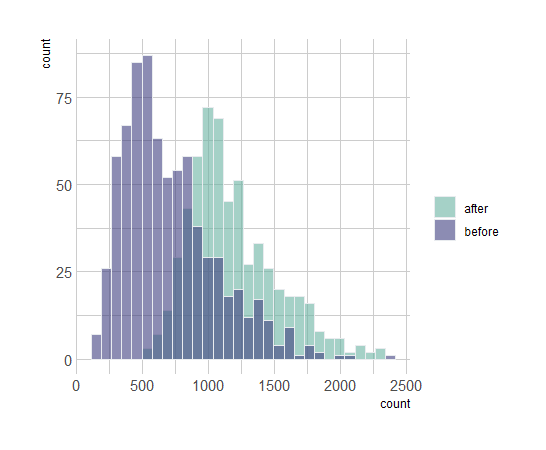}\\
\textbf{Figure 4}. Comparative histograms for the bird counts before 30/1/2020 and after.
\end{center}

\subsection{Phase wise comparison}

We will now try to outline the differences between bird counts by considering the phases in terms of several lockdowns which the government of India imposed, to curb the rising COVID cases. As discussed before, a belief is that due to the strict lockdowns, the number of sparrows would increase as a result of decreased human activities and all forms of pollution. We will focus on the five phases as described in Section 1, which are denoted as `before', `lock1', `after1', `lock2', and `after2' for convenience. A point to note is that the two lockdowns namely, first and second, coincided with the first and second COVID waves, which also had a global effect. Now the sample sizes (number of days) under these different phases won't be comparable, but we atleast aim to spot the differences visually. Table 2 contains some basic summary variables for the bird counts in different phases. One can note that during both the lockdowns, even though the sample size is much less than other phases, the bird counts are still comparable to other time periods. This indicates the human tendency of engaging themselves in other interests, such as bird watching, during strict lockdowns, when they were not allowed to even step out of their houses. For a better understanding, Figure 5 contains side-by-side boxplots and scatterplots of the sparrow counts for the different phases. These present similar conclusions as was seen from the summary table. An interesting observation from the scatterplots is that during summers of every year, the bird counts are seen to be the least. This may correspond to the fact that people prefer staying at homes during peak summers and hence amounting to reduced sparrow sightings.

\[%
\begin{tabular*}
{\textwidth}[c]{@{\extracolsep{\fill}}c|ccccccc}%
\multicolumn{8}{c}{\textbf{Table 2}. Summary variables of bird counts for each phase}\\ \\  \hline
Phase & No. of days & Min & $Q_1$ & Median & Mean & $Q_3$ & Max \\\hline \\
before & $754$ & $131$ & $449.8$ & $625.5$ & $710.9$ & $906.5$ & $2356$  \\
lock1 & $68$ & $879$ & $1246$ & $1506$ & $1469$ & $1654$ & $2220$  \\
after1 & $277$ & $589$ & $931$ & $1092$ & $1180$ & $1316$ & $2323$  \\
lock2 & $71$ & $865$ & $1014$ & $1090$ & $1142$ & $1234$ & $1584$ \\
after2 & $164$ & $544$ & $879.2$ & $1043$ & $1088.7$ & $1285$ & $2075$ \\ \hline \\
\end{tabular*}
\]

\bigskip

\[%
\begin{tabular*}
{\textwidth}[c]{@{\extracolsep{\fill}}cc}%
$\includegraphics[width = .45\textwidth]{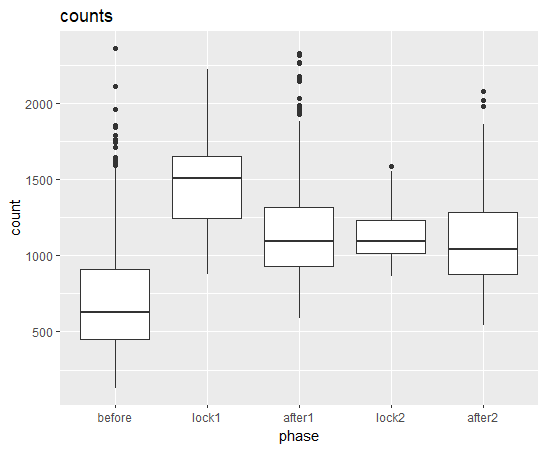}$ & $\includegraphics[width = .45\textwidth]{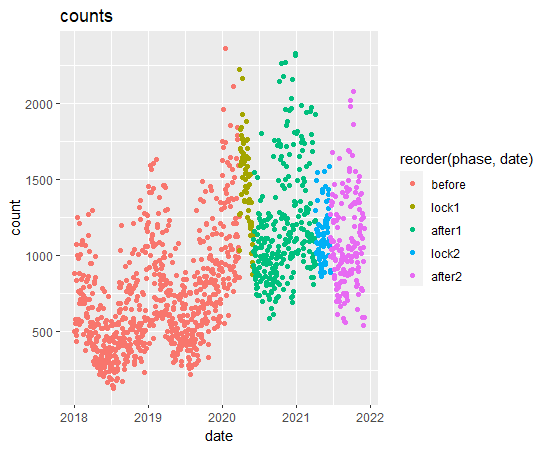}$\\
\multicolumn{2}{c}{\textbf{Figure 5}. Side-by-side boxplots (left) and scatterplots (right) for the different phases}
\end{tabular*}
\]

\bigskip

Another interesting comparison is that between sparrows sighted around sunrise (4am to 7am) and sunset (5pm to 7pm) on a given day. This will provide us with an idea as to how the sparrow sightings are distributed during the two most sought after time points for engaging in bird watching for any given day. In order to carry out this comparison, we adopted the following approaches. The first idea was to consider only the observations recorded around sunrise and sunset (separately) during the days of first lockdown in 2020 and the same dates in 2018. We purposefully picked 2018 as opposed to 2019 to ensure complete absence of any COVID traits. We observed that the mean daily bird counts at sunrise increased from 163.8 in 2018 to 597.8 in 2020 and that at sunset increased largely from 79.5 in 2018 to 620.2 in 2020. Figure 6 contains comparative scatterplots separately for sunrise and sunset, which show a connection between the sparrow counts for the same dates in 2018 and 2020 (during the first lockdown). One can note that during the first lockdown in 2020, more mass is towards the top of the vertical line, which indicates more number of higher bird count frequencies. The second idea was to understand the overall effects of COVID and lockdown. We thus adopted a similar technique which is seen before, to split the timeline at the date 30/01/2020, which amounts to `before COVID' and `after COVID' and compared the counts at sunrise and sunset. Table 3 contains some basic summary variables for this comparative study. One can clearly note that after the onset of COVID, the sparrow counts increased, both during sunrise and sunset.

\[%
\begin{tabular*}
{\textwidth}[c]{@{\extracolsep{\fill}}cc}%
$\includegraphics[width = .45\textwidth]{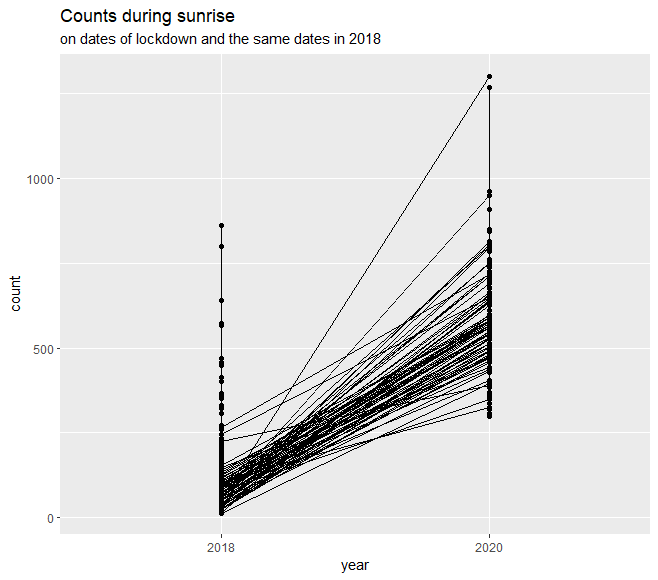}$ & $\includegraphics[width = .45\textwidth]{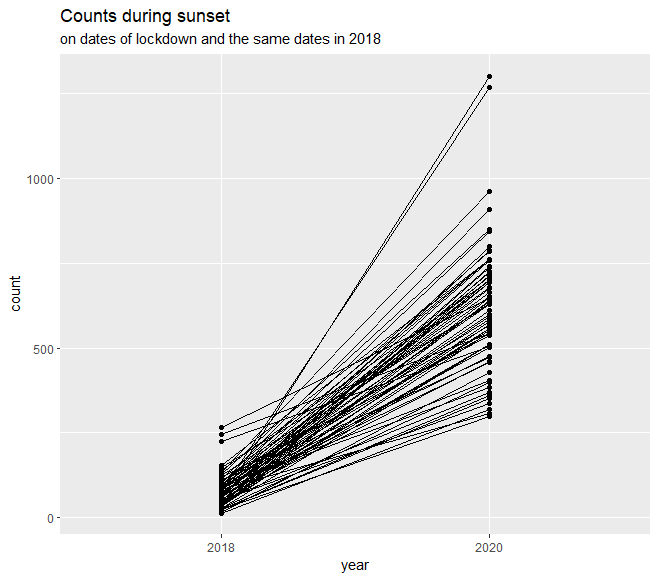}$\\
\multicolumn{2}{c}{\textbf{Figure 6}. Comparitive scatterplots of bird counts during sunrise (left) and sunset(right)}
\end{tabular*}
\]

\bigskip

\[%
\begin{tabular*}
{\textwidth}[c]{@{\extracolsep{\fill}}c|cccccc}%
\multicolumn{7}{c}{\textbf{Table 3}. Summary variables of bird counts during sunrise and sunset}\\ \\  \hline
Phase & Min & $Q_1$ & Median & Mean & $Q_3$ & Max \\\hline 
\multicolumn{7}{c}{\textbf{Sunrise}}\\ \cline{3-6}
before  & $8$ & $100.2$ & $167$ & $220.3$ & $277.2$ & $1386$  \\
after  & $29$ & $251$ & $361$ & $403.3$ & $519$ & $1341$  \\ 
\multicolumn{7}{c}{\textbf{Sunset}}\\ \cline{3-6}
before  & $8$ & $77$ & $116$ & $150.8$ & $177.8$ & $1386$  \\
after  & $29$ & $199$ & $267.5$ & $314.2$ & $367.5$ & $1326$ \\ \hline \\
\end{tabular*}
\]

\subsection{Statistical modelling}

Since the sparrow sightings is a count data, one would expect that a Poisson model would fit the observations better. In this section, we will try to implement suitable Poisson regression models, by taking a few related predictors and hence analysing their impact on the bird count. There have been similar attempts in the past to model the bird counts using a suitable regression technique. One may refer to Link et al. (2006), where the authors discuss about a hierarchical model to applied on Christmas bird counts, or C\c elik and Durmu\c s (2020), where the authors have statistically evaluated bird populations in Akdoğan Lakes by means of Poisson and negative binomial regression models. Now, if we focus on the raw data (without daily aggregation) it is found to contain more zeroes than expected. The excess zeroes could be generated by some other process than the Poisson process, like naturally low sparrow count of the location, lack of observer expertise, etc. Figure 7 contains a simple histogram for the bird counts coming from the raw data. One can clearly note the significant spike at $0$. In this scenario, a Zero Inflated Poisson regression (ZIP regression) would perform better, accounting for the excess zeroes. Firstly, we fitted a standard Poisson regression model to predict count using two variables namely, ``locality type" and ``starting hour" of observation. Here the variable ``locality type" refers to the code which is meant to help define the type of location used, as participants in eBird can plot specific locations on a map (P or Personal) or choose existing locations from a map (H or Hotspot). 

\begin{center}
\includegraphics[width = .55\textwidth]{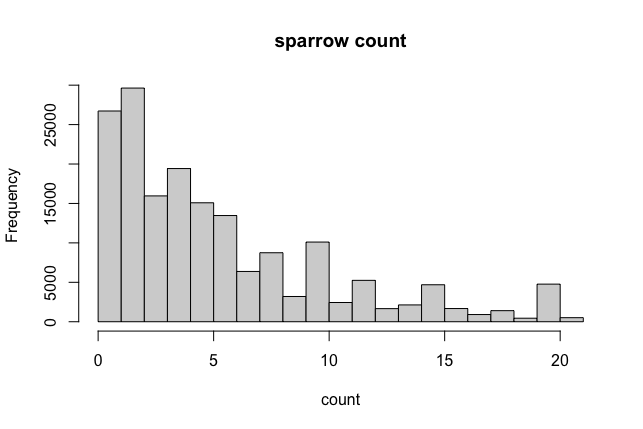}\\
\textbf{Figure 7}. A simple histogram of bird counts indicating excess zeroes
\end{center}

\bigskip

We then used a ZIP regression model using the same variables for predicting counts and phase of lockdown, for modelling the excess zeroes. A `Vuong' test was used to compare the two models. The Vuong test can be employed to compare predicted probabilities of non-nested models and is generally used in comparing zero-inflated models with standard models. If the two models were not different, we would see a large p-value. In our case, the p-value was small $(< 2.2 \times 10^{-16})$ which indicated that the ZIP model performed better than the standard Poisson one. ZIP regression is an example of the type of statistical model called the `latent variable model', where two or more processes occur together, which affect the response variable. In our case, we used the variables locality type and starting hour of observation for predicting counts and phase of lockdown for modelling the excess zeroes. There are two parts to this model. The first one is the standard Poisson regression part which is given by,

\[\log(\lambda)=\beta_0+\beta_1(\text{locality type})+\beta_2(\text{hour}),\]

while the other part models the excess zeroes and is given by,

\[ logit(\alpha)=\beta_0+\beta_1(\text{locality type})+\beta_2(\text{phase 2})+\beta_3(\text{phase 3})+\beta_4(\text{phase 4})+\beta_5(\text{phase 5}), \]

where the five phases are as taken in Section 2. Note that we have picked the base category to be `Phase 1' (before first lockdown). All the results are in comparison to this base category. All the coefficients of the model in both count and inflation parts were statistically significant. For interpretation, we exponentiate all the coefficients of the model. It is observed that for a given starting hour of observation, a personal locality is expected to cite 1.143 times birds than a hotspot. And for a given locality type, the bird count increased by a factor of 1.0012 with 1 hour increase in the starting hour of observation. 
The variable ‘phase’ was used to predict excess zeroes in the model. The log odds of observing a true zero in phase 2 is lesser by -0.54770 compared to phase 1. It is lesser for phases 3, 4 and 5 by -0.42907, -0.51192 and -0.44459 respectively, compared to phase 1. This means that the probability of observing true zero in phase 1, i.e., before lockdown is 0.0642 while it is 0.0381, 0.0428, 0.0395 and 0.0421 for the next 4 phases respectively. 

\section{Discussion}

From this analysis one can note that the house sparrow sightings per day have increased significantly after the imposition of lockdown. The increase in the daily number of bird counts could be mainly because more people might have started birdwatching during the lockdown as a new hobby or interest. Birdwatching might have helped them ease the hardships of COVID and the lockdown. This is supported by the fact that the count is 1.143 times more in a personal locality than in a hotspot. Many people who reported might have done backyard birdwatching from their homes.

Another conclusion is that the sparrows have been more visible due to quieter cities and towns with fewer human interactions. There have been various reports on the declining sparrow population in India, and several individuals and organizations have taken several efforts to bring awareness about the protection of sparrows. As a result, sparrows might have been reviving. However, we do not have enough evidence to claim this. Some other reports, like ``The State of India’s Birds 2020" by a group of research and conservation organizations, observe that the population of house sparrows has been relatively stable during the past 25 years. Also, the number of sparrows observed per citing did not significantly increase in the lockdown periods. 
From the comparison of sparrow counts at sunrise and sunset, which are the popular hours of the day for birdwatching, during lockdown in 2020 and the same dates in 2018, we can see a significant increase. Similar arguments can also explain this.

The probability of observing true zeroes decreased slightly in all the phases since lockdown. This also tells us that since lockdown, the number of people observing and the duration of observation have increased. More sparrows are spotted in personal localities, indicating that since there were restrictions on movement, people watched birds more from their homes rather than travelling to hotspots. We used the categorical variable `phase' to predict excess zeroes since the excess zeroes might be produced by people observing for a shorter duration, and lockdown phases might affect the duration of observation. The lockdown has also provided birdwatchers more free time to observe, the duration of observation might have increased, and as a result, zero counts should be lesser. The results supported this argument as the probability of observing excess zeroes decreased since the lockdown.

\section*{Data Availability}

All data used for the analysis in this paper are not publicly shareable, but can be directly obtained from \textit{https://ebird.org/data/download}, after making a suitable request to eBird.

\section*{References}

Basile, M., Francesco Russo, L., Giovanni Russo, V., Senese, A. and Bernardo, N. (2021). Birds seen and not seen during the COVID-19 pandemic: The impact of lockdown measures on citizen science bird observations, \textit{Biological Conservation}.\\

C\c elik, E. and Durmu\c s, A. (2020). Nonlinear regression applications in modelling over-dispersion of bird populations, \textit{Journal of Animal and Plant Sciences} 30(2).\\

Chaudhary, B. (2020). Home point study of birds and mammals diversity allied to humans in lockdown of COVID-19 at Bharatpur, Chitwan, Nepal, \textit{Open Journal of Ecology} 10(9): 612-631.\\

Deepalakshmi, S. and Salomi, A. A. (2019). Impact of urbanization on House sparrow (Passer domesticus) diversity from Erode and Namakkal districts, Tamilnadu, India, \textit{International Journal of Advanced Research in Biological Sciences)} 6(11).\\

\par eBird Basic Dataset. Version: EBD\_relNov-2021. Cornell Lab of Ornithology, Ithaca, New York. Nov 2021.\\

Ghosh, S., Kim, K. H. and Bhattacharya, R. (2010). A survey on house sparrow population decline at Bandel, West Bengal, India, \textit{Journal of the Korean Earth Science Society} 31(5): 448-453.\\

Gordo O., Brotons L., Herrando S. and Gargallo, G. (2021). Rapid behavioural response of urban birds to COVID-19 lockdown, \textit{Proceedings of the Royal Society B: Biological Sciences}.\\

Link, W. A., Sauer, J. R. and Niven, D. K. (2006). A hierarchical model for regional analysis of population change using Christmas Bird Count data, with application to the American Black Duck, \textit{Condor} 108: 13-24.\\

Modak, B. K. (2015). Impact of urbanization on house sparrow distribution: A case study from greater Kolkata,
India, \textit{Proceedings of the Zoological Society} 70: 21-27.\\

Narayanappa, Y., Gautam, A., Mahobiya, K. and Singh, A. (2022). Short Communication: A pilot-study on the occurrence and probable factors influencing the population decline of House Sparrow (Passer domesticus) along an urbanization gradient in Coimbatore district, India, \textit{Biodiversitas} 23(8).\\

Paul, M. R. (2015). A review of house sparrow population decline in India, \textit{Asia Pacific Journal of Research} 1(29).\\

Schrimpf, M. B., Des Brisay, P. G., Johnston, A., Smith, A. C., S\'anchez-Jasso, J., Robinson, B. G., Warrington, M. H., Mahony, N. A., Horn, A. G., Strimas-Mackey, M., Fahrig, L. and Koper, N. (2021). Reduced human activity during COVID-19 alters avian land use across North America, \textit{Science Advances} 7.\\

Seress, G., S\'andor, K., Vincze, E., Pipoly, I., Bukor, B., \'Agh, N. and Liker, A. (2021). Contrasting effects
of the COVID‑19 lockdown on urban birds’ reproductive success in two cities, \textit{Scientific Reports}.\\

Sharma, P. and Binner, M. (2020). The decline of population of house sparrow in India, \textit{International Journal of Agricultural Science} 5.

\end{document}